# Origin of Ultrafast Ag Radiotracer Diffusion in Shear Bands of Deformed Bulk Metallic Glass $Pd_{40}Ni_{40}P_{20}$

K. L. Ngai[1*], and Hai Bin Yu[2]

[1]*Dipartimento di Fisica, Università di Pisa, Largo B. Pontecorvo 3, I-56127, Pisa, Italy*

[2]*I. Physikalisches Institut, Universität Göttingen, Friedrich-Hund-Platz 1, D37077, Göttingen, Germany.*

**Abstract**

Measurements of Ag radiotracer diffusion in shear bands of deformed bulk metallic glass, $Pd_{40}Ni_{40}P_{20}$, [Joachim Bokeloh, Sergiy V. Divinski, Gerrit Reglitz, and Gerhard Wilde, Phys. Rev. Lett. **107**, 235503 (2011).] have found a colossal enhancement of diffusion coefficient by more than eight orders of magnitude than in undeformed $Pd_{40}Ni_{40}P_{20}$. Suggestion was made by Bokeloh et al. that enhanced diffusion occurs in high-mobility pathways originating from some excess free volume distribution inside the shear bands. Although plausible, this qualitative suggestion does not allow quantitative calculation of the enhancement. The impasse is avoided by using the Coupling Model to calculate the maximum of the enhancement of diffusivity possible in high-mobility pathways of the shear bands. Within the range of eight to ten orders of magnitude, the calculated maximum enhancement is capable to account for the experimental observation.

## I. Introduction

The scientific interest in bulk metallic glasses is prompted in part by their unusual mechanical properties with potential applications in areas of materials engineering. In some applications the bulk metallic glass (BMG) is subjected to high stress or strain, resulting in plastic deformation of the material. It is generally believed microscopically that the basic process underlying deformation is a local rearrangement of atoms. Argon [1,2] proposed the basic process involves few tens of atoms in a local cluster that changes from one low energy configuration to another by crossing energy barrier, which is now commonly referred to as a ''shear transformation zone'' (STZ). Spaepen [3,4] proposed an alternative mechanism of plastic deformation as discrete atomic jumps in the BMG near sites of larger free volume. The original model involving single-atom jumps had been revised in favor of multi-atom rearrangements, and the mechanism becomes more compatible with STZ. After the formation of the STZs and their self-organizations, macroscopic plastic deformation appears in very narrow regions called shear bands [5]. Transmission electron microscopy showed that the shear is sharply localized in shear band with thickness of 10 to 20 nm [6]. A comprehensive review of mechanical properties of BMGs was given in Ref.[7].

The atomic structure of the shear bands has been studied for the last thirty years by electron scattering [8], quantitative high-resolution transmission electron microscopy [9,10], and positron annihilation measurement [11]. Although differing in details, all these investigations had postulated that the shear bands in BMGs contain a higher concentration of nanometer-scale voids that resulted from coalescence of excess free volume after removal of the applied stress and stoppage of plastic flow. The size of the void-like defects was suggested to have diameter of approximately 1 nm.

While the experimental studies and computer simulation in the past mostly have been focused on static structural properties, dynamic properties of atoms in shear bands are practically unknown. From the presence of high shear strain within the shear band, we can infer that significant fraction of the atoms in the shear band undergo displacements, which should cause dilatation and enhanced mobility of atoms. This is supported by various experimental studies [12,13] and simulations [14,15]. For examples, from experiment it was concluded on a microscopic level that shear band propagation by plastic deformation is governed by the transformation of atomic-scale clusters such as the STZs, which are continuously activated by the applied high stress to populate the entire shear plane [12]. From the simulation of a deformed Lennard-Jones $Cu_{50}Zr_{50}$ model glass [14], a sizeable down shift by about 0.4 eV in the distribution of activation energies after plastic deformation was reported. Another simulation [15] found nearest-neighbor correlation in the activation of all STZs in the direction of the applied stress in inhomogeneous deformation.

Experimentally it is the recent measurement of atomic diffusion in a deformed BMG by Bokeloh et al. [16] that has shed more light on the dynamics. These authors performed for the first time diffusion measurements of $^{110m}$Ag radioisotope in the BMG, $Pd_{40}Ni_{40}P_{20}$, deformed to form shear bands. The $^{110m}$Ag radioisotope was chosen because the atomic radii of Ag and Pd are similar, and thus diffusion rate of the tracer measured can be taken as approximately that of Pd. The diffusion measurements were carried out at 473 K, more than 100 K below the glass transition temperature, $T_g$, equal to 585 K of the BMG determined by differential scanning calorimetry. At such low temperatures deep into the glassy state, Ag diffused negligibly in the undeformed glass. By extrapolating the bulk diffusion coefficients of data of Pd [17] and Au [18] in the undeformed $Pd_{40}Ni_{40}P_{20}$ down to lower temperatures, a value of $D_V=10^{-25}$ m$^2$/s is obtained as an estimate of the Ag volume diffusion coefficient $D_V$ at 473 K of the undeformed glass. For two sample deformed to different strains, the diffusion coefficient $D_{sb}$ measured at 473 K by Bokeloh et al. have much larger values of $1,13\times10^{-17}$ m$^2$/s and $1.96\times10^{-17}$ m$^2$/s. Thus the observed diffusion in deformed samples can be attributed entirely to diffusion in the shear bands.

Comparing $D_{sb}$ with $D_V$, the diffusion coefficient of Ag in the shear bands they found are more than 8 orders of magnitude higher than the volume diffusion coefficients measured previously for Pd [17] and Au [18]. This colossal enhancement of atomic diffusivity in shear bands that remains even after the external stress field has been removed is astounding. Perhaps given the known dilated static structure of shear bands, it is unsurprising that there is an enhancement of diffusivity than in the undeformed BMG. Notwithstanding, it is challenging to explain quantitatively, or at least semi-quantitatively, the enhancement is more than 8 orders of magnitude.

The objective of this paper is to provide a semi-quantitative explanation of the enhancement of atomic diffusivity in shear bands based on the known characteristics of the structural α-relaxation and the secondary β-relaxation processes in the undeformed BMG, and their relation to atomic diffusivity, potential energy barrier of STZ, crystallization, and mechanical properties including brittle to ductile transition. The plausible changes of these relaxation processes in shear bands of deformed BMG are shown to potentially account for experimentally observed enhancement of atomic diffusivity in shear bands by more than 8 orders of magnitude.

**II. Structural α-relaxation and secondary β-relaxation in BMG**

Viscosity, $\eta$, is related to the structural α-relaxation time, $\tau_\alpha$, by the Maxwell relation. $\eta = G_\infty \langle \tau_\alpha \rangle$, where $G_\infty$ is the high frequency shear modulus, and $\langle \tau_\alpha \rangle$ is the average of $\langle \tau_\alpha \rangle$ over the distribution. The Stokes-Einstein (SE) relation between $\eta$ and diffusion coefficient $D$ has been shown to be valid for the largest constituent Pd of $Pd_{43}Cu_{27}Ni_{10}P_{20}$, a BMG closely related to $Pd_{40}Ni_{40}P_{20}$. The Maxwell relation together with the SE relation engenders proportionality between $\langle \tau_\alpha \rangle$ and $D$ of undeformed $Pd_{40}Ni_{40}P_{20}$. Thus the change of $\langle \tau_\alpha \rangle$ in the shear bands formed by deforming the BMG quantitatively should be the same as the observed change of the diffusion coefficient from $D$ in the undeformed BMG to $D_{sb}$ in the shear bands. In other words, the problem originating from the experimental finding of enhanced $D_{sb}$ by Bokeloh et al. can be reformulated as follows: In shear bands formed after deformation, $\langle \tau_\alpha \rangle$ is shorter than $\langle \tau_\alpha \rangle$ in undeformed BMG by more than 8 orders of magnitude.

In general for glass-forming systems, the structural α-relaxation is a cooperative and dynamically heterogeneous process participated by multiple basic units. At times before the onset of the α-relaxation, there is a special kind of secondary or *β*-relaxation which involves motion of the entire basic unit. To distinguish them from the unimportant secondary relaxations involving motion of part of the molecule [19,20], they are often referred to as the Johari-Goldstein (JG) β-relaxation [21]. Although this JG *β*-relaxation is not cooperative, it exhibits properties mimicking the α-relaxation [22,23]. Its relaxation time $\tau_\beta$ bears strong connection to $\tau_\alpha$ in various dependences including temperature and pressure. The behavior of the JG *β*-relaxation from experiments and molecular dynamics simulations reveal that it is the precursor of the α-relaxation, and has fundamental importance when considering the dynamics of glass-forming systems. If it is truly the precursor of the α-relaxation, this special class of secondary relaxations has to be universally present in all glass-formers, because without which the omnipresent α-relaxation would not be realized. The universality of the JG β-relaxation is supported by finding it in different kinds of glass-formers with diverse chemical bonding and physical structures [19,20,22,23] including glass-forming metallic alloys. The various properties it shows all indicate that it is a local and independent process at its onset. This is intuitively obvious in some glass-formers where its most probable relaxation time $\tau_\beta$ is observed to be orders of magnitude shorter than $\tau_\alpha$ of the non-local many-body α-relaxation. The much smaller activation energy of $\tau_\beta$ in the glassy state compared with that of the α-relaxation is another indication. Simulations of polymers also have brought out this characteristic. The most extensive evidence comes from the approximate equality of the observed $\tau_\beta(T)$ and the primitive relaxation time $\tau_0(T)$ of the Coupling Model (CM) [22,24-26], and conceptually the primitive relaxation is a local relaxation and genuine precursor to the α-relaxation. Since we shall use the CM in helping to explain the enhanced atomic diffusion in shear bands, a summary of the essence is given below.

As mentioned already, similar to the JG β-relaxation in properties is the primitive (or independent) relaxation of the Coupling Model (CM), which is also not cooperative and strongly connected to the α-relaxation. The connection can be seen from the key equation of the CM,

$$\tau_\alpha = \left[ t_c^{-n} \tau_0 \right]^{1/(1-n)}, \qquad (1)$$

which relates the relaxation time, $\tau_0$, of the primitive relaxation to $\tau_\alpha$. In Eq.(1) $t_c$ is the temperature insensitive crossover time from primitive relaxation with time dependence, exp(-$t/\tau_0$), to cooperative α-relaxation with the Kohlrasuch stretched exponential time dependence,

$$\varphi(t) = \exp\left[ -(t/\tau_\alpha)^{1-n} \right] \qquad (2)$$

The fraction of unity, (1-$n$), in Eq.(1) is the same as the stretch exponent $\beta_K\equiv(1-n)$ in Eq.(2). In context of the CM, $n$ is the coupling parameter. The value of $t_c$ is about 1 to 2 ps for soft molecular glass-formers and polymers as determined by quasielastic neutron scattering experiments and molecular dynamics simulations [22]. It is shorter for metallic systems with $t_c \approx$ 0.2 ps from simulations [27,28], which is reasonable as inferred from the stronger metallic bonds compared with van der Waals interaction in soft matter, and is also exemplified by the much higher shear modulus of BMG than molecular glasses [29,30].

Verified before in many glass-formers [20,21,23,24] is the good correspondence between the $\tau_\beta(T)$ and $\tau_0(T)$, calculated by Eq.(1), with $\tau_\alpha(T)$ and (1-$n$) determined by fitting the time or frequency dependence of the α-relaxation with the Kohlrasuch function in Eq.(2) or its Fourier transform respectively. The approximate relation,

$$\tau_\beta(T)\approx\tau_0(T), \qquad (3)$$

will be used in conjunction with Eqs.(1) and (2) in the Section III to understand the origin of the enhanced diffusivity in shear bands.

There is only one secondary relaxation found in BMG, and it is the JG β-relaxation. The postulated local nature of the JG β-relaxation of BMG has been given support from synchrotron X-ray characterization of an ultra-quenched BMG by Liu et al. [31]. They show that the relaxation originates from a short-range collective rearrangement of atoms. At temperatures far below $T_g$, $\tau_\alpha$ becomes exceedingly long that the structural α-relaxation is not observable in the laboratory time scale even when extended to a few days in the diffusion experiments of Bokeloh et al.[16] By contrast, $\tau_\beta$ is shorter than $\tau_\alpha$ by orders of magnitude in the glassy state, and can be seen in isochronal mechanical relaxation experiments [32-34]. Moreover, the β-relaxation of BMG is likely relevant for the present problem in view of several recent findings by experiments of the correlation between the β-relaxation of BMG with properties directly related to plastic deformation. For example, Harmon et al. [32] proposed that isolated STZ transitions confined within the elastic matrix are associated with the $\beta$ process. Yu et al. later found that the potential energy barrier of STZ is nearly the same as the activation energy of the β-relaxation [33]. The β-relaxation of a BMG is closely correlated with the activation of the structural units of plastic deformations and global plasticity, and the brittle to ductile transition and the β-relaxation follow similar time-temperature dependence [34]. Hence we examine the β-relaxation and its relation to the structural α-relaxation of undeformed $Pd_{40}Ni_{40}P_{20}$ as well as in shear bands formed by deformation in the following section. As we shall see, the relations lead to an explanation of the colossal enhancement of diffusivity in shear bands than in the undeformed BMG.

**III. Explanation of the colossal enhancement of diffusivity in shear bands**

The observation of more than 8 orders of magnitude enhancement of atomic diffusion in the shear bands of the BMG, $Pd_{40}Ni_{40}P_{20}$ glass than in the untransformed matrix obviously point to much higher atomic mobility in shear bands. This can be inferred from the presence of void defects, excess free volume and less dense atomic structure of the shear bands [7-11] discussed before in the Introduction. If the excess free volume is relevant, its distribution inside the shear bands is not known. Bokeloh et al. [16] suggested that the excess free volume inside the shear bands is not homogeneously distributed, but tends to accumulate at or near the interfaces between unmodified matrix and shear band, and provides high-mobility pathways at the interfaces between shear bands and matrix. To arrive at this suggestion, they considered by analogy the extremely large diffusion coefficient at grain boundaries measured for 1 nm thin interface layers in polycrystalline Cu-Bi alloys [35]. The exact nature of the high-mobility pathways in the shear bands and the atomic structure giving rise to them cannot be ascertained at this time. Also not known quantitatively is magnitude of enhancement of diffusivity in the pathways. Hence $D_{sb}$ cannot be calculated or estimated from this static consideration. Nevertheless, we can proceed to give the largest enhancement of atomic diffusivity possible by turning our attention to $<\tau_\alpha>$ of the undeformed BMG and its change in the shear bands. This is made possible because of the proportionality between $<\tau_\alpha>$ and diffusion coefficient of the radiotracer Ag as discussed in the previous section.

**IIIa. Dynamics of Undeformed $Pd_{40}Ni_{40}P_{20}$**

To proceed, we have to characterize the dynamics of undeformed $Pd_{40}Ni_{40}P_{20}$ by determining its α-relaxation time $\tau_\alpha(T)$ and the β-relaxation time $\tau_\beta(T)$ in the glassy state. To do this, we first determine the Kohlrausch stretch exponent, $(1-n)$, by fitting the Fourier transform of Eq.(2) to the isothermal dynamic shear modulus data of of undeformed $Pd_{40}Ni_{40}P_{20}$ by Schröter et al. [36]. There is only four decades of frequency available for the measurement. Therefore only the data at the lowest measurement temperature of 591 K are considered because they capture more fully the α-loss peak, $G''$, and are less influenced on the high frequency flank of the loss peak by the contribution of the unresolved JG β-relaxation. The fit with $(1-n)=0.50$ is presented in Fig.1. The broadening on the high frequency flank of the $G''$ loss peak also could be due to contributions from the smaller atoms Ni and particularly P having higher mobility than Pd as shown by the breakdown of SE relation for the smaller atoms, while Pd obeys the SE relation [13]. With the value of $n=0.50$, $t_c=0.2$ ps, and assuming that $\tau_\alpha=200$ s at $T_g=585$ K determined by DSC [16], the value of $\tau_0(T_g)=6.3\times10^{-6}$ s is obtained via Eq.(1), which can be taken approximately as the value of $\tau_\beta(T_g)$ in view of Eq.(3). The activation energy $E_{A\beta}=130$ kJ/mol of $\tau_0(T) \approx \tau_\beta(T)$ in the glassy state has been determined by dynamic mechanical relaxation measurements [33]. From $\tau_0(T_g) = \tau_\beta(T_g) = 6.3 \times 10^{-6}$ s and $E_{A\beta} = 130$ kJ/mol, the Arrhenius temperature dependence of $\tau_0(T) \approx \tau_\beta(T)$ at temperatures below $T_g$ is completely determined, as shown by the solid line in Fig.2. Next, Eq.(1) allows $\tau_\alpha(T)$ to be calculated from $\tau_0(T) \approx \tau_\beta(T)$ again with $n=0.50$, $t_c=0.2$ ps, and the results are shown by the solid line on top in Fig.2. Naturally the temperature dependence of the calculated $\tau_\alpha(T)$ is Arrhenius and the activation energy of the α-relaxation, $E_{A\alpha} = 130/(1-n)$ kJ/mol, with $n=0.50$, and $\tau_\alpha(T_g) = 200$ s. There are a number of molecular and polymeric glass-formers for which experimental data are available for the Arrhenius temperature dependences of $\tau_\alpha(T)$ and $\tau_\beta(T)$ at temperatures below $T_g$ [37]. The relation between $\tau_\alpha(T)$ and $\tau_\beta(T)$ including the relation, $E_{A\alpha} = E_{A\beta}/(1-n)$, between their activation energies are in accord with Eq.(1). The proven successes in these soft-matter glass-formers of the relation between $\tau_\alpha(T)$ and $\tau_\beta(T)$ gives credence to the $\tau_\alpha(T)$ of undeformed glassy $Pd_{40}Ni_{40}P_{20}$ calculated from $\tau_\beta(T)$ in the same way.

Another source of information on the β-relaxation comes from the recent findings of Yu et al. [38]. They found that in multicomponent metallic glasses the β-relaxations are closely related to the self-diffusion of the smallest constituting atoms, which is P in the present case of $Pd_{40}Ni_{40}P_{20}$. Diffusion of the smallest constituting atoms takes place within the temperature and time regimes where the β-relaxations are activated, and the two processes have the similar activation energies. The hopping rate of P atoms in $Pd_{40}Ni_{40}P_{20}$ had been determined by the stimulated echo technique of nuclear magnetic resonance (NMR) before [39]. The NMR experiment had measured the jump rates of P over a range of temperatures in glassy $Pd_{40}Ni_{40}P_{20}$. The hopping rate $\Omega$(Hz) has the Arrhenius temperature dependence given by $\Omega = \Omega_o \exp(-E_{A,NMR}/RT)$, where $\Omega_o = 7.3 \times 10^{-13}$ s$^{-1}$ [39]. The activation energy of P diffusion, measured by NMR, $E_{A,NMR}=125$ kJ/mol, is indeed close to $E_{A,NMR}=126$ kJ/mol of β-relaxation. We converted the hopping rates to times, $\tau_{P,NMR}(T)$. Shown by the lower dashed line in Fig.2, $\tau_{P,NMR}(T)$ is an alternative estimate of $\tau_\beta(T)\approx\tau_0(T)$ of undeformed glassy $Pd_{40}Ni_{40}P_{20}$. Again we use Eq.(1) to calculate $\tau_\alpha(T)$ from $\tau_{P,NMR}(T)=\tau_0(T)$ again with $n=0.50$, $t_c=0.2$ ps, and the results are shown by the solid line on top in Fig.2.

### IIIb. Dynamics in Shear Bands of Deformed $Pd_{40}Ni_{40}P_{20}$

On decreasing temperature of supercooled liquids towards glass transition, the rapid increase of $\tau_\alpha(T)$ and viscosity $\eta(T)$ or decrease in diffusivity $D(T)$ are usually totally attributed to decrease in specific volume and entropy. The Coupling Model (CM) suggests many-body relaxation caused by intermolecular interactions (interatomic interactions in metallic alloys) have additional effect to slow down of the α-relaxation as exemplified by Eq.(1) where the coupling parameter $n$ can be considered as a measure of the strength of the interaction. An example to demonstrate this is given by a recent paper [40]. The dependence of $\tau_\alpha(T)$, $\eta(T)$, and $D(T)$ on temperature and density $\rho$ is combined into the product $\rho^\gamma/T$ variable, where $\gamma$ is a material specific parameter. It was shown that the β-relaxation time $\tau_\beta$ as well as $\tau_0$ is dependent on exactly the same $\rho^\gamma/T$ variable [40], i.e. $\tau_\beta(\rho^\gamma/T)\approx\tau_0(\rho^\gamma/T)$, and Eq.(1) holds in connecting $\tau_\alpha(\rho^\gamma/T)$ to $\tau_\beta(\rho^\gamma/T)\approx\tau_0(\rho^\gamma/T)$.

Relevance of the effect of many-body relaxation caused by intermolecular interactions as described by Eq.(1) can be uniquely identified from properties of supercooled liquids that are controlled by $n$ and observed at constant temperature and density [22]. In the glassy and iso-structural state and the change of density or free volume is unimportant, intermolecular interactions solely determines $\tau_\alpha(T)$. More extensive many-body relaxation manifested by larger $n$ engenders longer $\tau_\alpha$ and larger Arrhenius activation energy $E_{A\alpha}$, as can demonstrated explicitly by Eq.(1) together with $\tau_\beta(T)\approx\tau_0(T)$ when rewritten as

$$\tau_\alpha(T) = [\tau_0(T)/t_c]^n \tau_0(T) \approx [\tau_\beta(T)/t_c]^n \tau_\beta(T). \tag{4}$$

Since $t_c$ is much shorter than $\tau_\beta(T)\approx\tau_0(T)$, it can be easily understood from Eq.(4) that larger $n$ results in longer $\tau_\alpha(T)$ and larger separation of $\tau_\alpha(T)$ from $\tau_0(T)$ or $\tau_\beta(T)$. Conversely, if intermolecular interaction and $n$ can be reduced under some circumstance, the consequence is much shorter $\tau_\alpha(T)$ and smaller separation of $\tau_\alpha(T)$ from $\tau_0(T)$ or $\tau_\beta(T)$. Such circumstance is realized in dielectric and mechanical measurements of structural α-relaxation of ultrathin films of polymers, and nano-confined polymers, and molecular glass-former [41-43]. The nanometer size of the samples reduces the length-scale of the many-body relaxation and hence $n$, and it effect on slowing down $\tau_\alpha(T)$. Shorter $\tau_\alpha(T)$ means enhanced diffusion coefficient $D(T)$. Diffusion on the free surface of indomethacin (IMC), a molecular glass-former, had been measured by Zhu et al. [44]. On the free surface and in the absence of molecules on one side, the many-body effect on $\tau_\alpha(T)$ is much reduced. At temperatures at and below $T_g$, experimental data show the surface diffusion coefficient is enhanced by more than 6 orders of magnitude than that in the bulk. It was shown by applying the CM Eq.(4) to bulk IMC that at $T_g$ $\tau_\alpha(T)$ is also about 6 orders of magnitude longer than $\tau_\beta(T)\approx\tau_0(T)$ [45]. Thus on the surface, $\tau_\alpha(T)$ becomes about the same as $\tau_\beta(T)\approx\tau_0(T)$, indicating that $n$ is reduced to nearly zero [45].

Other examples not involving change of density and entropy include dynamics of ions in glassy and even crystalline ion conductors [46,47], where ion-ion interaction is the sole important determining factor of the many-ion dynamics. The primitive relaxation corresponds to independent single ion hop over energy barrier, and the relation between the measured ion conductivity relaxation time (the counterpart of $\tau_\alpha$) and the primitive ion hop relaxation time is governed by $n$ in the same manner as described by Eq.(1). We draw analogy of ion dynamics in glassy ionic conductors to the present case of hopping of Ag, Au or Pd atoms deep in the glassy state of undeformed $Pd_{40}Ni_{40}P_{20}$. Once more, the effect of atom-atom interactions embodied by $n$ is solely important in determining $\tau_\alpha(T)$, $\eta(T)$, and $D(T)$ of undeformed $Pd_{40}Ni_{40}P_{20}$ in the glassy state, and Eq.(1) holds as discussed in the previous section.

The above discussion on many-body dynamics of other systems is relevant to our present problem of the colossal enhancement of Ag radiotracer diffusion in some high mobility pathways of enhanced $D_{sb}$ in shear bands of deformed $Pd_{40}Ni_{40}P_{20}$ [16]. The colossal enhancement of the Ag radiotracer diffusion coefficient, or the corresponding large reduction of $\tau_\alpha(T)$, in shear bands of deformed $Pd_{40}Ni_{40}P_{20}$ compared with undeformed $Pd_{40}Ni_{40}P_{20}$, is explained in the context of the CM by a large reduction of the coupling parameter $n$ in Eqs.(1) and (4). The largest reduction of $\tau_\alpha(T)$ occurs if $n$ were reduced to zero in the high mobility pathways. In this case, $\tau_\alpha(T)$ becomes the same as $\tau_\beta(T) \approx \tau_0(T)$ of undeformed $Pd_{40}Ni_{40}P_{20}$. Therefore the maximum reduction of $\tau_\alpha(T)$ in the shear bands at 473 K can be obtained from the ratio $\tau_\alpha(T)/\tau_0(T) \approx \tau_\alpha(T)/\tau_\beta(T)$ of undeformed $Pd_{40}Ni_{40}P_{20}$ in Fig.2. Located with abscissa at 1000/473 $K^{-1}$, the dotted line connecting $\tau_\alpha(T)$ and $\tau_0(T)$ at $T$=473 K indicates the maximum reduction of α-relaxation time is about 10 orders of magnitude, and the maximum enhancement of diffusion coefficient $D_{sb}$ of Ag radiotracer in the shear bands is also 10 orders of magnitude. This theoretical upper bound of enhancement of $D_{sb}$ is not far from the experimental value of 8 decades, suggesting interaction of Ag atoms with other atoms is severely reduced in high mobility pathways in shear bands, and $n$ is reduced from the value of 0.50 in undeformed $Pd_{40}Ni_{40}P_{20}$ not completely to zero but a smaller value of about 0.18. If the P atom hopping diffusion time, $\tau_{P,NMR}(T)$, from NMR shown by the dashed line in Fig.2 is used as the alternative estimate of $\tau_\beta(T)$, the reduction of $\tau_\alpha(T)$ or the enhancement of $D_{sb}$ in the shear bands at $T$=473 K is about 8.5 decades, closer to the value estimated by Bokeloh et al. from experiments.

Rapid crystal growth in shear bands was observed in a deformed Al-rich metallic glass by G. Wilde and H. Rösner [48]. The effect was attributed to the marked increase of mobility due to excess volume present inside the shear bands, but it is also consistent with the large reduction of $\tau_\alpha(T)$ approaching $\tau_\beta(T) \approx \tau_0(T)$ in the high mobility pathways shown in the above. In undeformed BMG, crystal growth rate is negligible deep in the glassy state because $\tau_\alpha(T)$ is exceedingly long. Although the faster β-relaxation has the potential, it is not as effective for crystallization because of its local motion. As shown by Ichitsubo et al. [49], crystallization was negligible up to 75 h in undeformed $Pd_{42.5}Ni_{17.5}Cu_{30}P_{20}$ when the sample were annealed at 10 degrees below $T_g$=300 C, not as low as the measurement temperature of crystallization [48] and diffusion [16] in shear bands. Only when ultrasonic vibrations at frequencies resonating with the β-relaxation was applied can crystallization occurs by the accumulation of atomic jumps associated with the β-relaxation [49].

## IV. Discussion and Conclusion

The huge 8 decades in enhancement of Ag radiotracer diffusion coefficient $D_{sb}$ of deformed $Pd_{40}Ni_{40}P_{20}$ found by Bokeloh et al. [16] deep in the glassy state of the undeformed material signals the existence of high-mobility paths in the shear bands. This is consistent with the excess free volume created by deformation in the shear bands accumulating at or near the interfaces between unmodified matrix and shear band to give rise to the high mobility paths as suggested by Bokeloh et al. [16]. Because mechanical integrity is maintained in the deformed BMG, the high-mobility pathways cannot fill the full width of the shear bands and are instead very thin channels. Bokeloh et al. likened the situation here to the extremely high grain boundary diffusion coefficient measured for 1 nm thin interface layers in Cu-Bi alloys [35]. Such a scenario of a confined liquid film that has a width of a few monolayers would also give rise to stick-slip behaviour, as it was observed for shear band activation through acoustic emission spectroscopy [12,50]

Despite the advance, a method is needed that allows one to obtain a quantitative estimate of the enhanced $D_{sb}$ in the high-mobility pathways. In this paper we apply the Coupling Model to determine first the structural α-relaxation times $\tau_\alpha(T)$ of the undeformed $Pd_{40}Ni_{40}P_{20}$ at temperatures way below $T_g$ from the experimentally accessible β-relaxation time $\tau_\beta(T)$ or equivalently the primitive relaxation time $\tau_0(T)$. The maximum reduction of $\tau_\alpha(T)$ in the high mobility pathways in shear bands of deformed $Pd_{40}Ni_{40}P_{20}$ happens when many-body relaxation is reduced to become the local independent relaxation. The size of the maximum reduction can be obtained from $\log[\tau_\alpha(T)/\tau_\beta(T)]$, where both $\tau_\alpha(T)$ and $\tau_\beta(T)$ are for the undeformed BMG in this expression. By this procedure, we obtain $\log[D_{sb}/D_V]$ directly from $\log[\tau_\alpha(T)/\tau_\beta(T)]$, and the orders of magnitude of the maximum possible enhancement of diffusivity of Ag radiotracer in the high-mobility pathways in shear bands. Although the Coupling Model (CM) can only give an upper bound of the enhancement, its order of magnitude is comparable to the experimental result. The finding should be helpful in further experimental study of the physical structure of high mobility pathways in the shear bands.

It is worthwhile to recognize that the same method was applied successfully [45] to account for the 6 orders of magnitude enhancement of surface diffusion coefficient found in the molecular glass-former, indomethacin near $T_g$=315 K [44]. Due to total absence of molecules on one side of the free surface, it is easier to understand the reduction of intermolecular coupling for molecules diffusing exclusively on the surface. The vanishingly small coupling parameter $n$ is justified by the surface diffusion coefficient having about the same activation energy as $\tau_0(T) \approx \tau_\beta(T)$ [45]. Thus $n \approx 0$ for surface diffusion, and the enhancement of diffusivity on the surface compared to in the bulk is the same as the ratio of $\tau_\alpha(T)$ to $\tau_0(T) \approx \tau_\beta(T)$ in bulk indomethacin. The ratio is in accord with the 6 orders of magnitude enhancement of diffusivity found by experiment [44]. The reader may recognize that this is exactly the same way we obtained the upper bound of the enhanced $D_{sb}$ of Ag radiotracer diffusion in the high mobility pathways in shear bands.

There is yet another experiment on dynamics on surface of metallic glasses with results that support the treatment given in this paper. Observation of the dynamics of atoms on the surface of metallic glasses at temperature deep in the glassy state was made possible by analyzing scanning tunneling microscopy movies with time resolution as fast as 1 min and extending up to 1000 min [51,52]. The Fe-based metallic glass used has $T_g$=507 C, and the measurements were made at room temperature and in the range between 80 and 150 C. Surface clusters of atoms, all with compact structures with a width of 2-8 atomic spacings along the surface plane, rearrange themselves almost exclusively by two-state hopping back and forth by ~0.5 nm. The two-state dynamics of the surface clusters is spatially and temporally heterogeneous, and its rate is thermally activation with an average activation energy of $14k_BT$ and a preface of $(1 \text{ ps})^{-1}$. These properties are typical of the β-relaxation or primitive relaxation of metallic glasses. The smaller activation energy of $14k_BT$ compared with the typical $26k_BT$ of β-relaxation of BMGs is due to lower packing density of atoms at the surface than in the bulk. It is worthwhile to point out the measurement temperatures in this case is more than 350 degrees below $T_g$ in this case, while they ranges from a few degrees above $T_g$ to at most about 25 degrees below $T_g$ in the experiment on surface diffusion in indomethacin [44]. The difference in temperature range engenders large difference in the mobility of processes observed in the two experiments.

Reduction of intermolecular coupling occurs in glass-formers when confined in space or thin films of nanometer size without formation of chemical bonds with the substrate or walls. Like at free surface and in shear bands, the reduction leads to decrease of $\tau_\alpha(T)$. In fact when dimension of the confinement becomes sufficiently small, the surface layers dominate and the decrease of $\tau_\alpha(T)$ is so severe that it approaches $\tau_0(T) \approx \tau_\beta(T)$ [41-43]. In some experiments on ultrathin polymer films, the β-relaxation contributed by the surface can be seen directly [53-56].

**Acknowledgment**

Hai Bin Yu thanks the Alexander von Humboldt Foundation for support with a post-doctoral fellowship.

**Figure Captions**

Fig.1. Normalized shear loss modulus data of undeformed $Pd_{40}Ni_{40}P_{20}$ at 318 C (symbols) obtained by Schröter et al. [33] and fitted by the Fourier transform of the Kohkrausch function with stretch exponent $(1-n)$ equal to 0.50 (line).

Fig.2. The structural α-relaxation time $\tau_\alpha$ is assumed to be 200 s at $T_g$=585 K determined by DSC [13]. The primitive relaxation time at $T_g$, $\tau_0(T_g) \approx \tau_\beta(T_g)$, is calculated by Eq.(1) with $(1-n)$ equal to 0.50 determined in Fig.1. The lower solid line represents $\tau_0(T) \approx \tau_\beta(T)$ at lower temperatures using $\tau_0(T_g) \approx \tau_\beta(T_g)$ and the known activation energy of 130 kJ/mol of $\tau_\beta(T)$ from isochronal mechanical measurements of the β-relaxation. The upper solid line represent $\tau_\alpha(T)$ below $T_g$ calculated by Eq.(1) from $\tau_0(T_g) \approx \tau_\beta(T)$. The dashed line represents the P atom hopping diffusion time determined by NMR. The dotted line located at 1000/473 $K^{-1}$ is drawn to indicate 10 orders of magnitude difference between $\tau_\alpha(T)$ and $\tau_0(T) \approx \tau_\beta(T)$ at $T$=473 K.

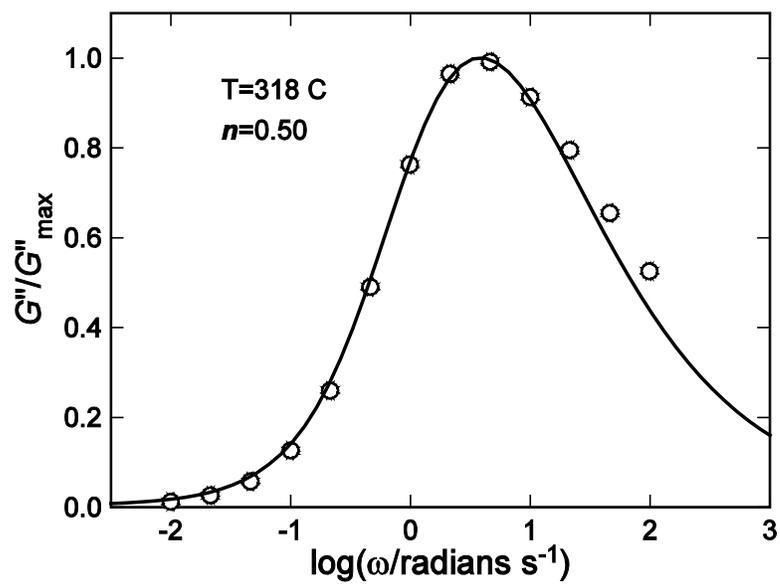

Figure 1

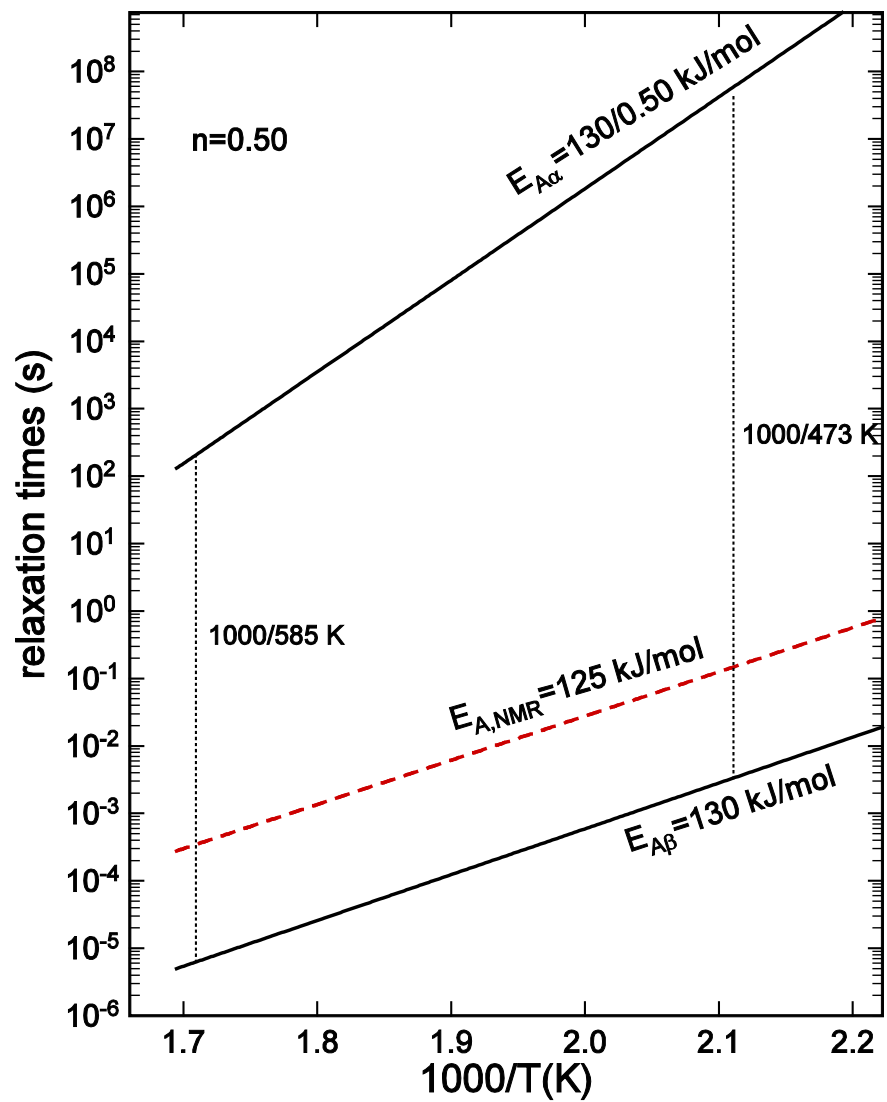

Figure 2.